\documentclass[journal]{IEEEtran}

\usepackage{cite}
\usepackage[dvips]{graphicx}
\graphicspath{{./}}
\DeclareGraphicsExtensions{.eps}
\usepackage{amsmath}
\usepackage{amsfonts}
\usepackage{array}
\usepackage[utf8]{inputenc}
\usepackage{multirow}
\usepackage[flushleft]{threeparttable}

\newcommand{\Z}{\mathbb{Z}}

\newcommand{\figureWidth}{0.9\columnwidth}
\newcommand{\plotWidth}{0.9\columnwidth}

\begin{document}

\title{EMC Regulations and Spectral Constraints for Multicarrier Modulation in PLC}

\author{Mauro~Girotto,~\IEEEmembership{Member,~IEEE,}
        and~Andrea~M.~Tonello,~\IEEEmembership{Senior Member,~IEEE}
\thanks{A version of this manuscript has been submitted to the IEEE Access for possible publication.\newline
        M. Girotto is with the University of Udine, Udine 33100, Italy (e-mail: mauro.girotto@uniud.it).
       A. Tonello is with the University of Klagenfurt, Klagenfurt 9020, Austria (e-mail: andrea.tonello@aau.at).}}


\maketitle

\begin{abstract}
This paper considers Electromagnetic Compatibility (EMC) aspects in the context of Power Line Communication (PLC) systems. It offers a complete overview of both narrow band PLC and broad band PLC EMC norms. How to interpret and translate such norms and measurement procedures into typical constraints used by designers of communication systems, is discussed. In particular, the constraints to the modulated signal spectrum are considered and the ability of pulse shaped OFDM (PS-OFDM), used in most of the PLC standards as IEEE P1901 and P1901.2, to fulfill them is analyzed. In addition, aiming to improve the spectrum management ability, a novel scheme named Pulse Shaped Cyclic Block Filtered Multitone modulation (PS-CB-FMT) is introduced and compared to PS-OFDM. It is shown that, PS-CB-FMT offers better ability to fulfill the norms which translates in higher system capacity.
\end{abstract}

\begin{IEEEkeywords}
Power line communication, EMC, regulations, multicarrier modulation, PS-OFDM, CB-FMT.
\end{IEEEkeywords}

\IEEEpeerreviewmaketitle

\section{Introduction}
\label{Sec:Introduction}
\IEEEPARstart{P}{ower-line} Communication (PLC) is an interesting and potentially cost effective solution to create a communication network exploiting the existing power delivery infrastructure. The application scenarios include home networking, home automation, broadband internet access, automatic meter reading, smart grid services through the power distribution grid \cite{PLC_state_art}.

PLC systems are partitioned in two classes: Narrow Band PLC (NB-PLC) and Broad Band PLC (BB-PLC). NB-PLC systems operate in the low frequencies range (below $500$ kHz) and offer low data-rate up to 500 kbps. NB-PLC is mostly intended for command and control and smart grid applications where reliability and robustness are preferred over data-rate. On the contrary, BB-PLC systems operate in the spectrum $1.8$--$30$ MHz, and more recently, up to 86 MHz. BB-PLC is mostly used for in-home networking where high data-rate communications, in excess of 200 Mbps, are required, e.g., for multimedia applications. 

The exact spectrum used by PLC depends on the continent or country where it is deployed. These band plans are defined by the local committees, e.g., CENELEC (Comité Européen de Normalisation Électrotechnique) in the EU, FCC (Federal Communications Commission) in the US and ARIB (Association of Radio Industries and Businesses) in Japan. For instance, CENELEC defines for NB-PLC four bands in the $3$--$148.5$ kHz spectrum \cite{EN50065}, while FCC allows NB-PLC in the $9$--$490$ kHz spectrum \cite{FCC_15} and ARIB in the $10$--$450$ kHz spectrum \cite{STD-T84}. The local committees also set the Electromagnetic Compatibility (EMC) regulations. A general electrical device emits unintentionally (or intentionally in the PLC case) signals in the electrical network that may cause interference to other devices. To prevent malfunctioning, EMC regulations define a set of emission and immunity limits that every electrical device must fulfill. For PLC, emission limits have a fundamental role since they introduce a constraint to the level of transmitted signal used for communications \cite[Chapter~2]{MIMO_PLC_Book}, \cite[Chapter~5]{BookPLC}. 

Usually, in communications, the involved signals and the system performance, e.g., the signal-to-noise ratio, the bit-error-rate and the capacity, are analyzed as a function of the transmitted signal Power Spectral Density (PSD). However, EMC regulations are rather elaborated and set the conducted and radiated emission limits following a precise measurement procedure. This procedure involves, for the conducted emissions, a spectrum analyzer with a certain Intermediate Frequency (IF) filter and a certain detector structure (EMI receiver), i.e., a peak, quasi-peak or average signal level detector. The conducted emission limits are expressed in dB$\mu$V. The results of such a procedure are not of immediate translation into the spectral constraints that the digital modulated signal has to fulfill.

Given, the spectral constraints and the multi-path characteristics of PLC channels, up-to-date NB-PLC and BB-PLC standards have adopted Multicarrier Modulation (MCM) \cite[Chapter~5]{BookPLC}. This is because in MCM, the informative signal is split into low data-rate signals. These signals are transmitted over narrow band sub-channels that can handle the channel frequency selectivity and enable simpler one-tap equalization. Furthermore, spectrum shaping can be implemented by switching off some sub-channels, i.e., through the creation of spectral notches. Spectrum notching is also required to allow coexistence with radio systems especially in the BB spectrum.   

Orthogonal Frequency Division Multiplexing (OFDM) \cite{Bingham} is the most common form of MCM. However, aiming to have a better spectrum confinement, PLC systems have opted for a modified version of it named Pulse Shaped OFDM (PS-OFDM) \cite{Pulse-Shaping}, \cite[Chapter~5]{BookPLC}. This is the case of the IEEE P1901.2 NB-PLC standard and of the IEEE P1901 BB-PLC standard.

In order to have a better spectrum management flexibility, other MCM techniques have been investigated as Wavelet OFDM \cite{Wavelet_OFDM}, OQAM/OFDM \cite{OQAM, OFDM_OQAM}, Filtered Multitone (FMT) \cite{FMT_Cherubini, FMT_Tonello_PLC}. More recently, a new Filter Bank Modulation (FBM) technique, named Cyclic Block Filtered Multitone Modulation (CB-FMT), that deploys cyclic convolutions instead of linear convolutions, has been introduced in \cite{Tonello_Girotto_JASP_2014} and analyzed for possible application to both NB-PLC in \cite{CBFMT_GLOBECOM} and to BB-PLC in \cite{CBFMT_ISPLC}. 
In CB-FMT, differently from OFDM, the sub-channel frequency confinement is preferred. This allows to reduce the out-of-band interference and to better realize notching and satisfy spectral masks. 

The contribution of this paper is twofold. Firstly, a  complete overview of relevant EMC norms for both NB-PLC and BB-PLC is reported. The procedure used to make the measurements and determine the limits is explained, as well as, it is discussed how to convert such limits into PSD constraints to the modulated signal. This has practical and theoretical relevance, since the topic is often debate in the PLC R\&D community.
Secondly, the implications of such limits into the design of PS-OFDM are assessed. In particular, the parameters used in IEEE P1901.2 and P1901 will be assumed to verify whether the norms are fulfilled. Then, in order to see whether improved spectrum usage can be done, we consider CB-FMT and compare it to PS-OFDM. In particular, we consider a new CB-FMT scheme that deploys not only frequency domain pulse shaping but also time domain pulse shaping. We refer to this scheme as Pulse Shaped CB-FMT (PS-CB-FMT).     
The analysis shows that PS-CB-FMT offers significant better fulfillment of norms both in the NB and in the BB spectrum. This is a result of the better ability to create notches and shape the spectrum. In turn, this implies higher capacity offered by the transmission scheme.    

This paper is organized as follows. In Sec.~\ref{Sec:EMC}, EMC regulations are analyzed in detail. The measurement procedures and limits are reported. In Sec.~\ref{Sec:Conversion}, the problem of the conversion from EMC limits to PSD limits for the modulated signal is addressed. In Sec.~\ref{Sec:PLC_standards}, the IEEE P1901.2 and IEEE P1901 standards are considered. Then, FBM is briefly explained and the novel PS-CB-FMT scheme is introduced. In Sec.~\ref{Sec:Results}, the maximum PSD limits are evaluated for both PS-OFDM and PS-CB-FMT. Furthermore, the maximum achievable rate offered by PS-OFDM and PS-CB-FMT is evaluated in typical outdoor low voltage and in-home scenarios. Finally, in Sec.~\ref{Sec:Conclusions}, the conclusions follow.

\section{Electromagnetic Compatibility Regulations}
\label{Sec:EMC}
In the context of EMC regulations and PLC, the ``Comité International Spécial des Perturbations Radioélectriques'' (CISPR) plays a fundamental role \footnote{The International Electrotechnical Commission (IEC) -- a non-profit, non-governmental international standards organization -- was founded in 1904 to standardize electrical apparatus and machinery. IEC established the birth of an international committee for defining EMI measurement methods and emission limits, namely CISPR, in 1933.\cite{EMC_Paul}}.
CISPR belongs to a non-governmental organization and it has not regulatory authority. The CISPR standards must be adopted by the local government to have legal effect. In the EU, the ``Comité Européen de Normalisation en Électronique et en Électrotechnique'' (CENELEC) starting from CISPR standards produces the European Norms (EN). These norms, so called harmonized standards, are then adopted by the EU members local committees, e.g., CEI in Italy, DKE in Germany, OVE in Austria. In the US, the CISPR standards have been adopted by the Federal Communications Commission (FCC) \cite{EMC_Engineering}.

Two specific relevant EMC standards that have an impact on PLC are CISPR 22 \cite{CISPR_22} and FCC Part 15 \cite{FCC_15}. These standards specify the EMI limits. FCC Part 15 was published in 1979 and it is valid for commercial products marketed in the US. CISPR 22 was published later in 1985 and it is intended for commercial products marketed outside the US. For the EU, CISPR 22 was adopted by CENELEC with the norm EN55022 \cite{EN55022}.

\subsection{Method of Measurement and Limits}
\label{Sec:EMC:Measurement}
The EMIs are partitioned in two classes: the conduced and the radiated emissions. Conducted emissions are related to the interference that is conducted through the power line. Radiated emissions are related to the electric and magnetic fields generated by the device and that propagate outside the power conductors. In the EU, EN55022 specifies that conducted emissions have to be measured in the frequency range between $150$ kHz and $30$ MHz, while radiated emissions have to be measured only above $30$ MHz. More recently, the norm EN50065 (specifically referred to NB-PLC) has included specifications below $150$ kHz, as discussed in the following. FCC in the US, instead, specifies that radiated emissions have to be measured also below $30$ MHz.

The goal of the conducted emissions test is to measure the interference injected into the power line. The measurement setup and methodology allows replicability and it is described in the CISPR 16-1 standard \cite{CISPR_16_1}. Since the emission level is related to the network impedance and it is affected by noise, a Line Impedance Stabilization Network (LISN) is inserted between the device under test and the network. The LISN sets a constant impedance ($50 \Omega)$ and it blocks the noise coming from the network. The measurement is then made with a Spectrum Analyzer (SA). In the SA, the RF signal is translated to the Intermediate Frequency (IF) and it is filtered with a bandpass filter, namely the IF filter. The bandwidth of the IF filter determines the resolution of the closely spaced frequency components. The envelope of the output signal contains the information about the conducted emission level. Therefore, after the  IF filter, an envelope detector is inserted whose structure depends on whether the Peak (PK), Quasi-peak (QP) or Average (AV) signal level has to be measured.  

\begin{figure}[t]
\centering
\includegraphics[width=\figureWidth]{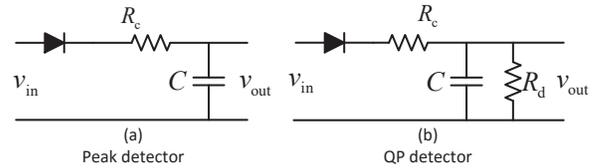}
   \caption{Peak and Quasi-Peak baseline schemes.}
   \label{Fig:detectors}
\end{figure}
\begin{figure}[t]
\centering
\includegraphics[width=\figureWidth]{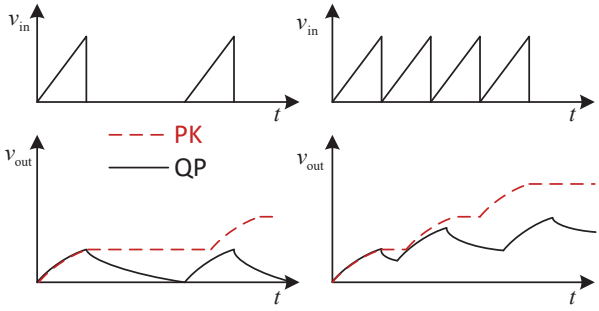}
   \caption{Qualitative output example of peak and quasi-peak detectors.}
   \label{Fig:pk_qp_signal}
\end{figure}

The PK and QP detectors schematic is reported in Fig.~\ref{Fig:detectors}. To understand the behavior of the PK and QP detectors, the example of Fig.~\ref{Fig:pk_qp_signal} is considered. At the detector input, a signal formed by a train of triangular pulses is considered. In both detectors, the capacitor $C$ is charged through the resistance $R_\text{c}$ until the pulse turns off. In the PK detector, the capacitor does not discharge itself so that  the following pulses charge the capacitor until the peak is reached. In the QP detector, the capacitor discharges itself through $R_\text{d}$ when the pulse is turned off. If the pulses duty cycle is sufficiently small, the capacitor will completely discharge itself. When the period decreases, the output signal will continue to increase to some limit determined by the PK value. In other words, if the signal at the QP detector input is infrequent, the output level will be significantly smaller than that measured by the PK detector. In conclusion, if the measured PK value does not exceed the regulatory limit, then the QP value will not exceed the limit too. Instead, if the PK value exceeds the regulatory limit, then the QP value must be evaluated. The detectors characteristics are described in the CISPR 16-1 standard \cite{CISPR_16_1}.

The CISPR 22 standard distinguishes among the mains terminals port and the telecommunication port. The mains terminal port refers to the power supply port. The telecommunication port refers to the data transmission port. Although for most of the communication devices the two ports are distinct, in PLC such ports physically coincide. Both Differential Mode (DM) and Common Mode (CM) limits are defined. Assuming three conductors, i.e., Phase (P), Neutral (N) and Protective Earth (PE) conductors, the DM is associated to the generated current that flows through the P conductor and returns through the N conductor. The CM is associated to the current that flows simultaneously through the P and the N conductors and returns through the PE conductor. The LISN allows to convert such currents into voltages that are finally measured. 


Both the mains port and the telecommunication port limits are defined in terms of CM which is responsible for radiated emissions. The telecommunication signal is injected in DM. However, the network (power grid in the case of PLC) due to intrinsic asymmetries may introduce a conversion from DM to CM. This is referred to as Longitudinal Conversion Loss (LCL). The CM limit is established using an appropriate LCL.

Finally, CISPR 22 splits the devices in two classes, A and B. Class A devices are used in commercial or industrial environment. Class B devices are used in residential, commercial or industrial environment. Class B limits are more stringent than Class A limits. The limits for the main and the telecommunication ports for Class A and B devices are reported, in terms of CM, in Tab.~\ref{Tab:conducted_limits}.

\renewcommand{\arraystretch}{1.4}
\begin{table*}[bt]
  \caption{Conducted emissions limits (Class A and B) defined in the CISPR 22/EN55022 standards.}
  \label{Tab:conducted_limits}
  \centering
 
  \begin{threeparttable}
  	
  	\begin{tabular}[width=\figureWidth]{|l|c|c|c|c|c|c|c|c|} \cline{2-9}
  	\multicolumn{1}{c|}{ } &  \multicolumn{4}{ c| }{Class A\tnote{a}} & \multicolumn{4}{ c| }{Class B\tnote{a}} \\ \cline{2-9}
 	\multicolumn{1}{c|}{ } & \multicolumn{2}{ c| }{Main port limits} & \multicolumn{2}{ c| }{Telecom. port limits} & \multicolumn{2}{ c| }{Main port limits} & \multicolumn{2}{ c| }{Telecom. port limits}\\ \hline
 	  \multicolumn{1}{|c|}{Frequency range} & QP & AV & QP & AV & QP & AV & QP & AV \\ \hline \hline
   	 $150$ kHz -- $500$ kHz & 79 & 66  &  $97$ -- $87$\tnote{b}  & $84$ -- $74$\tnote{b} & $66$ -- $56$\tnote{b} & $56$ -- $46$\tnote{b} & $84$ -- $74$\tnote{b} & $74$ -- $64$\tnote{b} \\ \hline
  	  $500$ kHz -- $5$ MHz & \multirow{2}{*}{73} & \multirow{2}{*}{60} & \multirow{2}{*}{87} & \multirow{2}{*}{74} & $56$ & $46$ & \multirow{2}{*}{74} & \multirow{2}{*}{64} \\ \cline{1-1} \cline{6-7}
    	$5$ MHz -- $30$ MHz &  &  & &  & $60$ & $50$ & & \\ \hline

  \end{tabular}
  
  \begin{tablenotes}
     	\item[a] the limits are expressed in dB$\mu$V.
        \item[b] decreasing linearly with log. of frequency.
    \end{tablenotes}
 \end{threeparttable}
 
\end{table*}

\subsection{PLC Regulations}
\label{Sec:EMC:PLC}
The early versions of the CISPR 22 standard did not consider the special case of PLC where the main port for power supply and the telecommunication port coincide. To clarify the situation, the CISPR/I subcommittee was created. Two projects were devoted to PLC issues: the project ``CISPR amendment for PLC products'' (active from 1999 to 2005) and the project ``Amendment of the CISPR 22 with appropriate limits and methods for PLT devices'' (active from 2005 to 2010). In ten years of work, several reports were produced. The amendment CISPR/I/89/CD \cite{CISPR_I_89} proposed in 2003 to apply the telecommunication limits to PLC modems. This amendment was then used since 2003 to test the PLC modems by industry. Furthermore, in 2009, the amendment CISPR/I/301/CD \cite{CISPR_I_301} was proposed to introduce EMI mitigation techniques based on notching, similarly to the notching mask used in the HomePlug commercial modems \cite{HomePlug_mask}. Nevertheless, both projects failed without coming to a conclusive standard. For these reasons, the local committees have become active to find a solution. In the following, the status of PLC regulations in the EU and US is summarized.

\renewcommand{\arraystretch}{1.2}
\begin{table}[bt]
  \caption{EU In-Band and Out-of-Band limits for PLC according to EN50065 and EN50561-1.}
  \label{Tab:EU_limits}
  \centering
 
  \begin{threeparttable}
  	
  	\begin{tabular}{|l|c|c|c|} \cline{2-4}
 		\multicolumn{1}{c|}{ } & \multicolumn{3}{ c| }{In-band limits\tnote{a}} \\ \hline
   		\multicolumn{1}{|c|}{Frequency range} & Peak & Quasi-Peak & Average \\ \hline \hline
    	$3$ kHz -- $9$ kHz & 134 &  &  \\ \hline
    	$9$ kHz -- $95$ kHz (NB)\tnote{c} & $134$ -- $120$\tnote{b} &  &  \\ \hline
    	$9$ kHz -- $95$ kHz (WB)\tnote{d} & $134$ &  &  \\ \hline
    	$95$ kHz -- $148.5$ kHz & 122/134\tnote{e} &  &  \\ \hline
    	$150$ kHz -- $500$ kHz\tnote{f} &  & $115$ -- $105$\tnote{b} & $105$ -- $95$\tnote{b} \\ \hline
    	$1.6065$ MHz -- $30$ MHz & 105 &  & 95 \\ \hline
    	\multicolumn{4}{c}{  } \\ [1ex] \cline{2-4}
    	\multicolumn{1}{c|}{ } & \multicolumn{3}{ c| }{Out-band limits\tnote{a}} \\ \hline
    	$3$ kHz -- $9$ kHz & 89 &  &  \\ \hline
    	$9$ kHz -- $150$ kHz &  & $89$ -- $66$\tnote{b} &  \\ \hline
    	$150$ kHz -- $30$ MHz & \multicolumn{3}{ c| }{EN55022 main port limits (Tab.~\ref{Tab:conducted_limits})}  \\ \hline
	  \end{tabular}
	  
  	\begin{tablenotes}
  		\item[a] the limits are expressed in dB$\mu$V.
     	\item[b] decreasing linearly with log. of frequency.
        \item[c] Signals with bandwidth less than 5 kHz (NB).
        \item[d] Signals with bandwidth greater than 5 kHz (WB).
        \item[e] Class 122 devices for general use. Class 134 devices for the industrial environment.
        \item[f] Unofficial limits proposed in IEEE P1901.2 standard.
    \end{tablenotes}
 \end{threeparttable}
 
\end{table}

\subsubsection{European Union}
\label{Sec:EMC:PLC:EU}
In the EU, the relevant norms approved by CENELEC for PLC are: EN50065 for NB-PLC in the $3$--$148.5$ kHz band, and EN50561-1 \cite{EN50561_1} for BB-PLC in the $1.6065$--$30$ MHz band. The in-band and out-of-band limits adopted in the EU are summarized in Tab.~\ref{Tab:EU_limits}.   

The EN50065 norm regulates NB PLC spectrum by partitioning it in four sub-bands. In the early versions of EN50065, these bands were referred to as CENELEC A, B, C, D. In recent versions, this designation has been removed. However, for convenience, the bands are referred with the old names. CENELEC A ($3$--$95$ kHz) is reserved to utilities. CENELEC B ($95$--$125$ kHz), C ($125$--$140$ kHz) and D ($140$--$148.5$ kHz) are used for home, commercial and industrial applications. CENELEC C requires the use of the CSMA/CA access protocol. Based on this, currently in the EU the NB spectrum $148.5$--$500$ kHz is not regulated. In Tab.~\ref{Tab:EU_limits}, we report the unofficial limits proposed by the IEEE P1901.2 NB-PLC standard. 

In the EN50561-1 norm \cite{EN50561_1}, approved in 2012, the emission test is divided into two parts. First, the emissions are measured at the PLC port without any communication established. In this case, the CISPR 22 approach is followed. Then, the emissions are measured when the communication takes place. The emission level is measured in terms of DM voltage only, so that CM emissions (LCL approach of CISPR 22) are not measured. Furthermore, EN50561-1 defines a notching mask to protect certain parts of the radio spectrum. 

It is interesting to note that EN50561-1 applies only to indoor BB-PLC so that outdoor BB-PLC is currently unregulated. Despite the effort, the norm EN-50561-2 that intended to regulate outdoor BB-PLC in the EU was not approved. Very recently, in 2016, CENELEC has approved the norm EN50561-3 that regulates indoor BB-PLC in the spectrum $30-87.5$ MHz. 

\subsubsection{United States}
\label{Sec:EMC:PLC:US}
In the US, PLC limits are regulated by FCC with the Code of Federal Regulations, Title 47, Part 15 (47 CFR §15). 

In CFR §15.3(t), NB-PLC is referred to as ``power line carrier systems''. NB-PLC can operate in the spectrum $9$--$490$ kHz and used by power utilities. NB-PLC is not subject to conducted or radiated emission limits. CFR FCC Part 15 (§15.113 simply specifies that the devices shall be operated with the minimum power possible. If an interference issue is reported, the electric power utility shall adjust its power line carrier operation. However, these regulations do not apply to ``electric lines which connect the distribution substation to the customer or house wiring'' (§15.113(f)), e.g., those used for meter reading. For this scenario, the limits related to ``carrier current system'' in CFR §15.3(f) have to be considered. In particular, CFR §15.107(c) states that the only limits are for the out-of-band conducted emission so that the 535–1705 kHz band is protected. Furthermore, the paragraphs §15.109(e) and §15.209(a), specify in-band radiated emission limits for the band $9$--$490$ kHz.

BB-PLC according to FCC is referred to as ``Broadband over Power Line'' (BPL). BPL can operate in the frequency range $1.705$--$80$ MHz. Differently from the EU, where PLC is regulated only in the in-home scenario, in the US, BB-PLC is regulated both for the outdoor (Access BPL, defined in §15.601) and for the indoor (In-Home BPL, defined in §15.3(gg)) scenarios. Furthermore, radiated emissions are evaluated below $30$ MHz too \cite{MIMO_PLC_Book}. For BB-PLC between $1.705$ MHz and $30$ MHz the radiated field strength is set to $30$ $\mu$V/m at a distance of 30 m. For the frequency range $30$--$80$ MHz the limit is set to $100$ $\mu$V/m at a distance of 3 m. The radiated limits are reported in §15.209.

\section{Limits Computation and Conversion}
\label{Sec:Conversion}
As discussed, Tab.~\ref{Tab:EU_limits} reports the transmission limits for PLC in the EU. The question is how to interpret and verify that a certain modulated communication signal fulfills such PK, QP and AV values. Furthermore, it is interesting to see how those limits can be translated into PSD values since very often digital modulation schemes are designed so that a certain PSD is imposed. The answer to these questions is not trivial. It is required to first consider a specific IF filter and then model the behavior of the PK, QP and AV detectors. This is discussed in the following two sections. Then, starting from the IEEE P1901.2 and P1901 main parameters, we evaluate the spectrum of the PS-OFDM modulated signal, we show how it fulfills the limits and we determine the maximum PSD value that PS-OFDM can assume. Aiming to show that better spectrum usage can be accomplished yet fulfilling the norms, we consider the novel PS-CB-FMT scheme and compare it with PS-OFDM.      

\subsection{Ideal IF Filter}
\label{Sec:Conversion:Ideal}
As described in Sec.~\ref{Sec:EMC:Measurement}, the PK, QP and AV emission values are computed at the IF filter output. Therefore, the measured value depends on the specific IF filter used. Assuming an ideal IF filter, i.e., a rectangular shaped pass band filter, and a modulated signal with a flat PSD in the IF band, the mean PSD that the signal must possess is obtained from the level at the output of a PK detector, namely $V_\text{PK}(f)$, as follows \cite[Chapter~6]{MIMO_PLC_Book}
\begin{align}
\overline{\text{PSD}}(f) =& \left[  V_{dB \mu V}(f)-10 \log_\text{10} \left(2 Z_0 \right) + \right. \notag \\
& \qquad \qquad \left.  - 10 \log_\text{10} \left(B_\text{IF} \right) \right] - \nu \quad \left[ \frac{\text{dBm}}{\text{Hz}} \right], \label{eq:PSD_conversion}
\end{align}
where $V_{dB \mu V}(f)$ is the $V_\text{PK}(f)$ voltage expressed in dB$\mu$V, i.e., $V_{dB \mu V}(f) = 10 \log_{10} \left( V_\text{PK}\left(f \right)/10^{-6} \right)^2$. $Z_0$ and $B_\text{IF}$ are the measured signal spectral density in dB$\mu$V, the SA impedance ($50 \, \Omega$), and the IF filter bandwidth, respectively. According to CISPR 16-1, $B_\text{IF} = 220$ Hz for the frequency range between $9$ kHz and $150$ kHz; $B_\text{IF} = 9$ kHz for the frequency range between $150$ kHz and $30$ MHz. In \eqref{eq:PSD_conversion}, the term enclosed into the square brackets is expressed in dBpW/Hz. The $\nu$ coefficient allows the conversion between dBpW/Hz and dBm/Hz, i.e., $\nu = 90$. Since, in general, the telecommunication signals exhibit cyclostationary behavior \cite{cyclostationarity}, \eqref{eq:PSD_conversion} gives the mean PSD.  

From \eqref{eq:PSD_conversion}, the maximum PSD level in the considered range of frequencies that the transmitted signal can take is then obtained as 
\begin{align}
\text{PSD}_\text{lim} &=  \underset{f}{\arg \max} \left\{\overline{\text{PSD}}(f)\right\} \notag \\
 &= V_\text{lim}-10 \log_\text{10} \left(2 Z_0 \right) - 10 \log_\text{10} \left(B_\text{IF} \right) - \nu. \label{eq:Teo_conversion}
\end{align}
where $V_\text{lim}$ is the maximum level of the signal spectral density in dB$\mu$V, i.e., $V_\text{lim}=\underset{f}{\arg \max} \left\{  V_{dB \mu V}(f) \right\}$. 

As an example, considering BB-PLC, according to Tab.~\ref{Tab:EU_limits}, if we set $V_\text{lim} = V_\text{lim,AV}=95$ dB$\mu$V we will obtain the PSD limit $\text{PSD}_\text{lim} = -55$ dBm/Hz, which is the value adopted in most of the PLC literature.

\subsection{Limits Practical Computation}
\label{Sec:Conversion:RealConversion}
The limits reported in Tab.~\ref{Tab:EU_limits}, are expressed in terms of the PK, QP and AV signal values in dB$\mu$V. The modulated signal $x(t)$ has to be filtered by the IF filter,  then the filter output passes to an envelope detector and finally either to the PK, or the QP or the AV detector, so that it can be verified whether it fulfills the limit. The first issue is to specify the IF filter. CISPR 16-1 does not specify it. Rather, it provides upper and lower limits that its frequency response has to fulfill as shown in Fig.~\ref{Fig:GaussianIF} for both the $9$--$150$ kHz band ($B_\text{IF}$ = 220 Hz) and the $0.15$--$30$ MHz band ($B_\text{IF}$ = 9 kHz). Then, the signal detectors have to be modeled to allow a numerical evaluation on whether the modulated signal used by the PLC system fulfills the norms. Or in other words, the signal detectors have to be modeled to determine how to scale and shape the modulated signal so that norms are fulfilled. This is discussed in the next two sections assuming a digital implementation (as in state-of-the-art signal/spectrum analyzers) with sampling period $T_s$.

\subsubsection{IF Filter}
\label{Sec:Conversion:RealConversion:IF}
\begin{figure}[tb]
\centering
\includegraphics[width=\plotWidth]{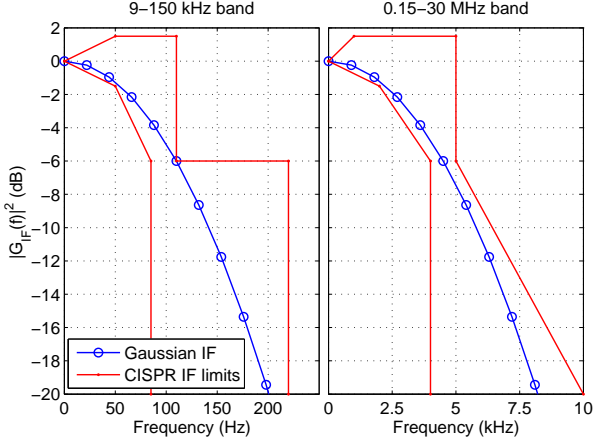}
   \caption{Frequency responses of Gaussian IF filter for for the $9$--$150$ kHz band (on the left) and $0.15$--$30$ MHz band (on the right). The plots include also the CISPR 16-1 bounds.}
   \label{Fig:GaussianIF}
\end{figure}
a simple filter that satisfies the CISPR 16-1 spectral bounds is the Gaussian filter. The impulse response is given by
\begin{equation}
g_\text{IF}(n T_s) = \frac{\sqrt{\pi}}{T_0} e^{-\pi^2  (n T_s/T_0)^2}, \label{eq:ifgaussian} 
\end{equation}
where $T_0$ is a constant that sets the IF filter bandwidth and $T_s$ is the sampling frequency of the digital SA. CISPR 16-1 specifies the 6 dB bandwidth. In this case, $1/T_0^2 = 5/6 \cdot B_\text{IF}^2 \ln 10$. Fig.~\ref{Fig:GaussianIF} shows the Gaussian frequency response and the fulfillment of the CISPR 16-1 bounds.

\subsubsection{PK Detector}
\label{Sec:Conversion:RealConversion:PK}
the PK detector measures the maximum value of the IF signal. It can be digitally implemented as
\begin{equation}
V_\text{PK} = \underset{0 \leq nT_s \leq T_\text{m}}{\arg \max}\left\{x_\text{ev}(nT_s)\right\},
\end{equation}
where $x_\text{ev}(nT_s)$ is the envelope of the IF filter output and $T_\text{m}$ is the measurement time.

\subsubsection{QP Detector}
\label{Sec:Conversion:RealConversion:QP}
to model the QP (non-linear) detector, the approach described in \cite{QP_digital} is followed. The analog QP detector, shown in Fig.~\ref{Fig:detectors}, is a typical $RC$ circuit with different charge ($\tau_c$) and discharge ($\tau_d$) time constants. The time constants values are reported in Tab.~\ref{Tab:QP_time_constant}.
\renewcommand{\arraystretch}{1.2} 
\begin{table}[bt]
  \caption{QP charge and discharge time constants and IF bandwidths.}
  \label{Tab:QP_time_constant}
  \centering
  	\begin{tabular}{|l|cc|c|} \hline 
     Band & $\tau_c$ (ms) & $\tau_d$ (ms) & $B_\text{IF}$ (kHz) \\ \hline \hline
     $9$--$150$ kHz & 45 & 500 & 0.22 \\ \hline
     $0.15$--$30$ MHz & 1 & 160 & 9 \\ \hline
   
  \end{tabular}
  
\end{table} 
If the input signal $v_\text{in}(t)$ is greater than $v_\text{out}(t)$ (capacitor voltage), then the capacitor will be charged through $R_c$. Oppositely, if the input signal $v_\text{in}(t)$ is less than $v_\text{out}(t)$, then the capacitor will be discharged through $R_d$. Since $\tau_d \gg \tau_c$, the QP detector can be modeled as two distinct $RC$ circuits, with Laplace transfer function given by
\begin{equation}
H(s) = \frac{1}{1+\tau s}, \quad \tau = \begin{cases}
    \tau_c & \text{for } v_\text{in}(t) \le v_\text{out}(t) \\
    \tau_d & \text{otherwise}
  \end{cases}. \label{eq:qp_tf}
\end{equation}

The analog transfer function \eqref{eq:qp_tf} can be translated into the discrete time domain by applying the bilinear transform defined as \cite{oppenheim1999discrete} 
\begin{equation}
s = \frac{2}{T_s} \frac{z-1}{z+1}. \label{eq:bilinear}
\end{equation}
After some algebraic manipulation, the Z-transform of the $RC$ filter is given by
\begin{equation}
 H_d(z) = \frac{b_0+b_1 z^{-1}}{1 + a_1 z^{-1}}, \label{eq:qp_tf_discrete}
  \end{equation}
  where
  \begin{align}
 b_0 = \begin{cases}
    0 & \text{discharge}\\
    b_1 & \text{charge}
  \end{cases},& \quad b_1 = \frac{1}{1+2\tau/T_s} \notag\\
  a_1 =& \frac{1-2\tau/T_s}{1+2\tau/T_s} \notag
  \end{align} 
The transfer function in \eqref{eq:qp_tf_discrete} represents an Infinite Impulse Response (IIR) filter. Consequently, the digital QP output signal is given by
\begin{equation}
 V_\text{QP}(n T_s) = b_0 x_\text{ev}(n T_s) + b_1 x_\text{IF}((n-1)T_s) - a_1 V_\text{QP}((n-1)T_s),
 \end{equation} 
where $b_0 = b_1$ for $x_\text{ev}(n T_s) \le  V_\text{QP}((n-1)T_s)$ and $0$ otherwise.

The digital QP detector herein described fulfills the CISPR 16-1 requirements.

\subsubsection{AV Detector}
\label{Sec:Conversion:RealConversion:AV}
the AV detector measures the mean value of the IF signal. It can be digitally implemented as
\begin{equation}
V_\text{AV} = \frac{1}{T_m} \sum_{n=0}^{T_m/T_s-1} x_\text{ev}(nT_s)
\end{equation}


\subsubsection{Implications to Signal PSD}
\label{Sec:Conversion:RealConversion:STOPSD}
based on the above procedures, it can be verified whether the designed modulated signal fulfills all PK, QP and AV limits imposed by norms. Then, it is possible to deduce what the  specific signal PSD is. In particular, the PSD in dBm/Hz is computed from \eqref{eq:PSD_conversion} and \eqref{eq:Teo_conversion}.


\section{PLC Standards and Adopted Modulation Schemes}
\label{Sec:PLC_standards}
The up-to-date PLC standards have adopted multicarrier modulation in the form of PS-OFDM. This has been done on one side to be able to handle the severe multipath propagation in PLC channels and on the other side to enable a flexible usage of the spectrum to fulfill the EMC norms. In this section, we briefly report a summary of the standards and, focusing on IEEE ones, we summarize the relevant parameters. Furthermore, we show that PS-OFDM and PS-CB-FMT, can be obtained via a cyclic filter bank. The spectral properties of both systems are then investigated.

\subsection{IEEE P1901.2 Standard}
\label{Sec:PLC_standards:NB}
Recently, several NB-PLC systems based on OFDM have been introduced to increase rate and to improve the spectrum usage flexibility, i.e., PRIME \cite{PRIME} and G3-PLC \cite{G3}. These technologies exhibited some coexistence issues. To solve the problem, in 2010, the "International Telecommunication Union" (ITU) started the G.hnem project. The goal was to produce a unified standard for NB-PLC \cite{Ghnem_intro}. In 2012, the standard ITU G.9902 (known as G.hnem) \cite{GHNEM} was released, which essentially incorporated features of both PRIME and G3. 

With similar scope to the G.hnem project, in 2009 IEEE started the P1901.2 working group which delivered the IEEE P1901.2  standard in 2014 \cite{P1901_2}. The P1901.2 inherits several characteristics of PRIME and G3 but with increased spectrum usage flexibility to cover the spectrum up to $500$ kHz and with band plans that enable the operation in the EU, the US and Japan. 

In all the systems mentioned, and in particular in IEEE P1901.2, PS-OFDM has been adopted as multicarrier modulation scheme. The details of this modulation scheme are briefly reported in the next section. Essentially, the number of sub-carriers is set to $K=128$. The sampling frequency depends on the operating band. In CENELEC bands the sampling frequency is set equal to $400$ kHz. In FCC band, the sampling frequency is set equal to $1.2$ MHz. The band plan of IEEE 1901.2 for the EU and the US is reported in Tab.~\ref{Tab:1901_bandplan}.

\renewcommand{\arraystretch}{1.2}
\begin{table}[bt]
  \caption{Band plan of IEEE P1901.2 for the EU and US.}
  \label{Tab:1901_bandplan}
  \centering

  \begin{threeparttable}
  	\begin{tabular}{|c|l|ccc|}\cline{3-5}
  	 \multicolumn{2}{c|}{  } & \multicolumn{3}{c|}{ Carriers } \\ \cline{2-5}
    \multicolumn{1}{c|}{  }	  & \multicolumn{1}{c|}{Band} & First\tnote{a} & Last\tnote{a} & $N_\text{ON}$\tnote{b}\\ \hline
    	\multirow{2}{*}{EU} &  CENELEC A & 35.9 & 90.6 & 36\\
     	&  CENELEC B & 98.4 & 121.8 & 16\\ \hline
    \multirow{2}{*}{US}   &	FCC-above-CENELEC & 154.6 & 487.5 & 72\\
       &	FCC-Low\tnote{c} & 37.5 & 117.2 & 18\\ \hline
   
  \end{tabular}
  
  \begin{tablenotes}
  		\item[a] Carriers are expressed in kHz.
  		\item[b] Number of active carriers.
     	\item[c] FCC-Low is used jointly with FCC-above-CENELEC.
    \end{tablenotes}
 \end{threeparttable}
 
\end{table}  

\subsection{IEEE P1901 Standard}
\label{Sec:PLC_standards:BB}
For BB-PLC, two standards have been released. The G.hn standard has been released in 2009 \cite{GHN} and the IEEE P1901 standard has been released in 2010 \cite{P1901}. In particular, the IEEE P1901 standard reports specifications for both indoor and outdoor scenarios. At the physical layer, PS-OFDM has been adopted in one transmission interface. There is also another multicarrier option which is named Wavelet OFDM \cite{Wavelet_OFDM}. In the PS-OFDM interface, the number of sub-carriers is set to $K=2048$ to cover the band $1.8$--$28$ MHz. The sampling frequency is set equal to $100$ MHz.

\subsection{From Standard PS-OFDM to PS-CB-FMT}
\label{Sec:PLC_standards:FBM}
To explain how PS-OFDM (used in most of the standards) works, we firstly introduce the expression of a general multicarrier modulated signal 
\begin{equation}
x(nT) = \sum_{k=0}^{K-1} \sum_{\ell \in \mathbb{Z}} a^{(k)}\left(\ell N T \right) g \left(nT-\ell N T \right) W_K^{-nk}, \label{eq:FBM_tx}
\end{equation}
where $K$ and $T$ are the sub-channels number and the sampling period, respectively. $a^{(k)}\left(\ell N T \right)$ are the data symbols belonging to the QAM signal set that are transmitted in the $k$-th sub-channel with symbol period $NT$. The overall data rate is equal to $K/(NT)$. $g(nT)$ is the prototype pulse impulse response and $W_K^{-nk} = e^{i 2 \pi nk/K}$. In the following, the sampling period is assumed to be normalized to simplify the notation, i.e., $T=1$. The prototype pulse is deployed to shape the sub-channels. In OFDM, it corresponds to a rectangular pulse of duration $K$ samples, or $N>K$ samples if the cyclic prefix (CP) is used. In Filtered Multitone Modulation (FMT), the prototype pulse is designed so that the frequency confinement is privileged \cite[Chapter~5]{BookPLC}. In PS-OFDM the prototype pulse is a smoothed window, e.g., a raised-cosine window in time domain.   

The modulated signal \eqref{eq:FBM_tx}, after digital-to-analog conversion, is injected and transmitted over the PLC channel. The received signal, after analog-to-digital conversion, is given by
\begin{equation}
y(n) = x * g_\text{ch}(n) + \eta(n), \label{eq:FBM_rx_input} 
\end{equation}
where $g_\text{ch}(n)$ and $\eta(n)$ are the channel impulse response and the background noise, respectively. Finally, the signal is demodulated with a bank of analysis filters with prototype pulse $h(n)$ and sampled to obtain 
\begin{equation}
z^{(k)}(m N) = \sum_{n \in \mathbb{Z}} y(n) W_K^{nk} g(mN -n), \label{eq:FBM_rx}
\end{equation}
with $k \in \left\{ 0, \dots, M-1 \right\}$. The analysis prototype pulse is equal to a rectangular pulse of duration $K$ samples in OFDM and CP-OFDM, while it is matched to $g(n)$ in FMT. Furthermore, sub-channel equalization is implemented to detect the data symbols.

As discussed, PS-OFDM is in most cases deployed in PLC. In the next section, we show that it can be derived starting from a different multicarrier architecture formulation that uses cyclic convolutions and not linear convolutions to synthesize and analyze the multicarrier signal. 

To proceed, we replace the linear convolution in \eqref{eq:FBM_tx} and \eqref{eq:FBM_rx} with a circular convolution, so that we will obtain a new form of multicarrier modulation that we refer to as  Cyclic Block Filtered Multitone Modulation (CB-FMT) \cite{Tonello_Girotto_JASP_2014}. CB-FMT is a filter bank modulation architecture that exploits cyclic filter banks \cite{CFBM}. It should be also noted that blocks of data are processed by the cyclic filter bank. The CB-FMT modulated signal related to the $q$-th block is given by
\begin{align}
x_q(n) &= \sum_{k=0}^{K-1} \sum_{\ell=0}^{L-1} a^{(k)}(q L + \ell N) g \left( \left(n- \ell N \right)_M \right) W_K^{-nk},\\
  n & \in \left\{ 0, \dots, M-1 \right\}, \notag 
\end{align}
where $(n)_M$ denotes the modulo $M$ operation on the integer coefficient $n$, so that $g((n)_M)$ is the periodic repetition of the prototype pulse, i.e., $g((n+aM)_M) = g(n)$, $\forall a \in \mathbb{Z}$, and $M = LN$. The block $q$ of data symbols $a^{(k)}(q L + \ell N)$ carries $L$ data symbols per sub-channel $k$ yielding a total of $KL$ data symbols per block.  

OFDM can still be obtained from this formulation by setting $K=N$, $L=1$ and $g(n)=1$ for $n \in \left\{ 0, \dots, K-1 \right\}$. More in general, CB-FMT uses a certain prototype pulse $g(n)$ to obtain a better sub-channel frequency confinement than the sinc-like shape obtained with OFDM. Both orthogonal and non-orthogonal filter bank designs can be realized. The implementation of CB-FMT can be done using a concatenation of Discrete Fourier Transforms (DFTs) \cite{Tonello_Girotto_JASP_2014}. The overall complexity is significantly lower than in filter bank modulation based on linear convolution. 

A Cyclic Prefix (CP) can also be added to each block. The CP will transform the linear convolution of \eqref{eq:FBM_rx_input} in a circular convolution if the CP duration is greater than the channel impulse response duration. Consequently, equalization can be performed easily in the frequency domain. The CP addition extends the length of the blocks from $M$ to $M + \mu$ coefficients.  

\subsection{PS-OFDM, PS-CB-FMT and Spectral Properties}
To analyze the spectral properties of the considered modulation schemes, the discrete-time blocks of coefficients must be concatenated and interpolated to obtain a continuous time signal. The concatenation of the blocks of coefficients can be further followed by (time domain) pulse shaping. This pulse shape, denoted with $g_\text{ps}(n)$, is a window  of $M + \mu + 2\alpha$ coefficients, e.g., a raised cosine window. In the specific case of OFDM, this generates the so called pulse shaped OFDM solution. In the case of CB-FMT, the procedure generates Pulse Shaped CB-FMT (PS-CB-FMT).   

Consequently, the pulse shaped signal can be written as 
\begin{equation}
x(n) = \sum_{q \in \Z} x_q((n-qM_1-\mu-\alpha)_M) g_\text{ps}(n-qM_1). \label{eq:x_cont_1}
\end{equation}
with $M_1= M + \mu + \alpha$.
It should be noted that in conventional CP-OFDM the pulse shaping filter has a rectangular response, i.e., $g_\text{ps}(n)=1$ for $n \in \{0, \dots, M_1-1\}$ and 0 otherwise. Since its frequency response has a sinc-like spectrum, the Out-of-Band (OOB) emissions are high and even more exacerbated by the sinc sub-channel frequency responses. Thus, the usage of pulse shaping enables to obtain a more compact spectrum and to ease the fulfillment of spectral norms. This can be shown analytically by computing the mean PSD of the PS-CB-FMT signal \eqref{eq:x_cont_1}. The result is given by 
\begin{align}
\overline{\text{PSD}}(f) = G_I(f) \frac{L}{M_1} & \sum_{k=0}^{K-1} \sum_{p=0}^{Q-1}  |G(p)|^2 |G_\text{ps}(f-f_{k,p})|^2, \label{eq:PSD}\\
Q = M/K, & \quad f_{k,p} = (p+kQ)/M, \notag
\end{align}
where $G(p)$ and $G_\text{ps}(f)$ are the $M$-point DFT of the prototype pulse, and the Fourier transform of the pulse shaping filter, respectively. Furthermore, we also consider the digital-to-analog converter through its frequency response $G_I(f)$. In the particular case of PS-OFDM, \eqref{eq:x_cont_1} is still valid by setting $Q=1$ and $G(0)=1$. In PS-CB-FMT, the adopted prototype pulse with DFT coefficients $G(p)$, intrinsically contributes to render the spectrum more compact w.r.t. PS-OFDM. 

\begin{figure}[tb]
\centering
\includegraphics[width=\figureWidth]{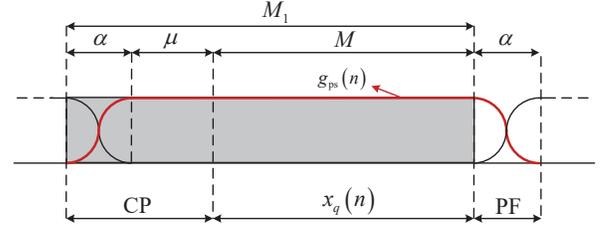}
   \caption{Raised cosine pulse shaping in PS-CB-FMT.}
   \label{Fig:PS}
\end{figure}
Now, PS-OFDM can deploy a raised cosine window with a roll-off of $\alpha$ samples. Similarly, in PS-CB-FMT we can use the following window
\begin{align}
g_\text{ps}(n) &= \begin{cases}
g_\text{rc}(n) & \mbox{for } n \in [0, \alpha] \cup [M_1,M_1+\alpha]\\
1 & \mbox{for } n \in ]\alpha, M_1 [\\
0 & \mbox{otherwise}
\end{cases}, \label{eq:PSRC}\\
g_\text{rc}(n) &= \frac{1}{2}+\frac{1}{2}\cos\left(\frac{\pi}{\alpha}\left(|n-n_1|-n_2\right)\right),\\
n_1 &= (M_1+\alpha)/2, \, n_2 = n_1 - \alpha, \, M_1 = M+\mu+\alpha. \notag
\end{align}
Since the PS filter in \eqref{eq:PSRC} has a length of $M_1+\alpha$ samples and the block period is set to $M_1$ samples, a Cyclic Postfix (PF) of $\alpha$ samples is added, and an overlap-and-add operation is implemented between adjacent blocks, as shown in Fig.~\ref{Fig:PS}. This overlap among blocks is introduced to limit the rate drop due to the PS, so that the transmission rate is equal to $R=LK/(M_1T)$ symb/s. 

The receiver of PS-CB-FMT performs firstly the removal of the CP to obtain $y_q(n)$. Then for each sub-channel, it runs a cyclic convolution between the received $q$-th block of coefficients and the analysis prototype pulse matched to the synthesis pulse, i.e., $h(n) = g^*(-n)$, to obtain
\begin{align}
z^{(k)}_q(mN) &= \sum_{n=0}^{M-1} y_q(n) W_K^{nk} g^* \left( \left(mN + n \right)_M \right), \label{eq:CBFMT_rx}\\
k & \in \left\{0, \dots, K-1 \right\}, \quad m \in \left\{0, \dots, L-1 \right\}. \notag
\end{align}

The cyclic analysis filter bank in \eqref{eq:CBFMT_rx} can be implemented in frequency domain using a concatenation of DFTs as shown in \cite{Tonello_Girotto_JASP_2014}. Minimum-mean-square-error equalization is also directly implemented in the frequency domain for each sub-channel $k \in \left\{0, \dots, K-1 \right\}$.

\begin{figure}[tb]
\centering
\includegraphics[width=\plotWidth]{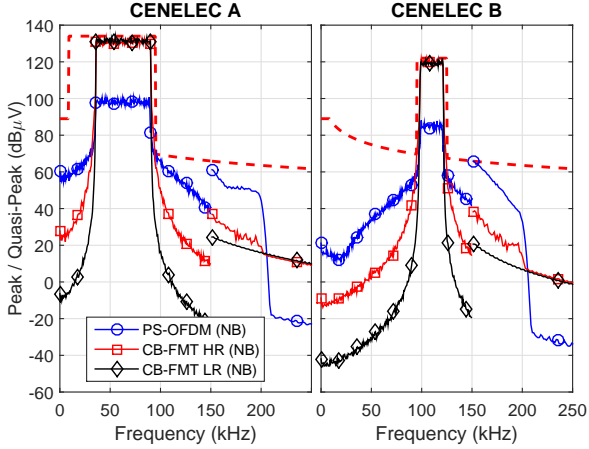}
   \caption{Peak/Quasi-Peak signal level for PS-OFDM, PS-CB-FMT and limit (CENELEC bands).}
   \label{Fig:QP_CENELEC}
\end{figure} 
\section{Fulfillment of EMC Norms}
\label{Sec:Results}
In this section, the aim is to verify that the PS-OFDM solution adopted in the IEEE PLC standards fulfills the EMC norms and to do so by using the methodology described in Sec.~\ref{Sec:EMC}. Furthermore, the novel PS-CB-FMT solution is considered and compared with the PS-OFDM solution to see whether a better fulfillment of the norms can be obtained. A comparison in terms of maximum achievable rate is also made.    
\subsection{System Parameters}
\label{Sec:Results:Params}
For the comparison, both PS-OFDM and PS-CB-FMT adopt the main specifications of IEEE P1901.2 and P1901 so that they use the same band plan, see also Tab.~\ref{Tab:1901_bandplan}. The overall parameters are listed in Tab.~\ref{Tab:CBFMT_params}. The ones for PS-OFDM have been taken from the standard. In PS-CB-FMT the prototype pulse has a rectangular frequency response. Time domain pulse shaping uses a raised cosine window. Furthermore, two configurations are considered:
\begin{enumerate}
\item PS-CB-FMT (referred to as High Rate (HR) configuration) with the same number of sub-channels $K$, CP length $\mu$, and roll-off $\alpha=\left\{8, 496\right\}$ as PS-OFDM, giving a peak normalized rate equal to $R=0.98$ for NB-PLC and $R=0.9$ for BB-PLC ; 
\item PS-CB-FMT as above but with $\alpha=\left\{239, 1264\right\}$ so that the rate is equal to $R=0.9$ for NB-PLC and $R=0.82$ for BB-PLC, which is referred to as Low Rate (LR) configuration.
\end{enumerate}

\begin{figure}[tb]
\centering
\includegraphics[width=\plotWidth]{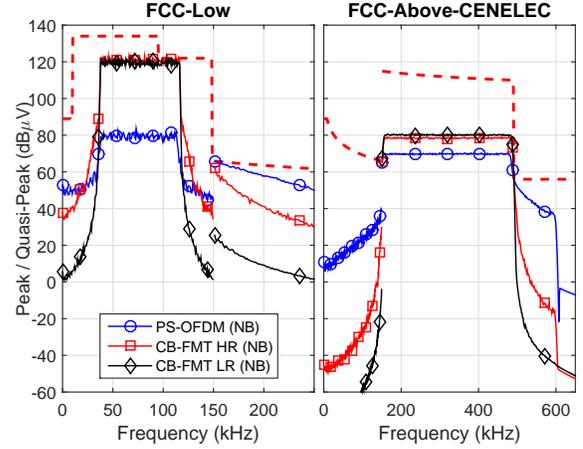}
   \caption{Peak/Quasi-Peak signal level for PS-OFDM, PS-CB-FMT and limit (FCC bands).}
   \label{Fig:QP_FCC}
\end{figure}
\renewcommand{\arraystretch}{1.4}
\begin{table*}[bt]
  \caption{PS-OFDM and PS-CB-FMT parameters.}
  \label{Tab:CBFMT_params}
  \centering
 
  	\begin{tabular}{|c|l|c|c|c|c|c|l|} \cline{3-8}
    \multicolumn{2}{c|}{ }  & $K$ & $N$ & $L$ & $\alpha$ & $\mu$ &  Acronym \\ \hline \hline
   \multirow{3}{*}{\rotatebox[origin=c]{90}{NB-PLC}} & PS-OFDM &  128 & 128 & 1 & 8 & 30 & PS-OFDM (NB)\\  \cline{2-8}
   &  \multirow{2}{*}{PS-CB-FMT} & 128 & 128 & 20 & 8  & 30 & PS-CB-FMT HR (NB) \\  \cline{3-8}
     &   & 128 & 128 & 20 & 239  & 30 & PS-CB-FMT LR (NB)\\ \cline{1-1} \hline \hline
   \multirow{3}{*}{\rotatebox[origin=c]{90}{BB-PLC}} & PS-OFDM &  2048 & 2048 & 1 & 496 & 556 & PS-OFDM (BB)\\  \cline{2-8}
   &  \multirow{2}{*}{PS-CB-FMT} & 2048 & 2048 & 4 & 496 & 556 &  CB-FMT HR (BB) \\  \cline{3-8}
     &   & 2048 & 2048 & 4 & 1264  & 556 & PS-CB-FMT LR (BB)\\ \cline{1-1} \hline

  \end{tabular}
  


\end{table*} 

%
%
%
  



\subsection{Spectral Limits}
\label{Sec:Results:Limits}
\subsubsection{NB-PLC}
since the NB-PLC EMC limits are specified in term of PK and QP values, the approach introduced in Sec.~\ref{Sec:Conversion:RealConversion} is adopted. Firstly, the modulated signal is generated. The transmission is assumed continuous and the data symbols are randomly chosen. Then, the signal is interpolated with an ideal interpolation filter.


Secondly, the PK and QP values are computed according to the methods discussed in Sec.~\ref{Sec:EMC:Measurement} and the signal is scaled so that the spectral limits are fulfilled. This means that signal level and the switched off sub-channels have to be adjusted so that the PK or QP values do not exceed both the In-Band (IB) and the Out-of-Band (OOB) limits. According to Tab.~\ref{Tab:EU_limits}, the IB limits in the EU are in terms of PK value while the OOB limits are both in terms of PK and QP values. For the FCC spectrum beyond the CENELEC bands, we assume the values in Tab.~\ref{Tab:EU_limits} as also reported in IEEE P1901.2. The IB limits are then in terms of QP value, while the OOB limits are in terms of PK or QP values.    

Figs.~\ref{Fig:QP_CENELEC} and \ref{Fig:QP_FCC} report the PK/QP values for both PS-OFDM and PS-CB-FMT for the CENELEC and FCC bands, respectively. 
The figures show a discontinuity at 150 kHz. This discontinuity is caused from the different measurement IF bandwidth for $f < 150$ kHz ($220$ Hz) and $f > 150$ kHz ($9$ kHz).

The figures show that PS-OFDM according to the configuration of P1901.2 fulfills the EMC limits. However, to meet the OOB limits the signal amplitude has to be lowered, so that the IB signal level is much below what would be in principle possible to use. On the contrary, PS-CB-FMT has much better spectrum confinement that allows to more efficiently fulfill the spectral masks. Therefore, the OOB limits can be met by a significantly higher IB signal level.  

The above considerations can also be done by looking at the maximum value of the PSD (PSD limit) that the signal can assume yet respecting the norms. This has been done, by computing the PSD limit for the considered modulated signals in Fig.~\ref{Fig:QP_CENELEC} and Fig.~\ref{Fig:QP_FCC}. The result is reported in Tab.~\ref{Tab:PSD_limit}. 
\renewcommand{\arraystretch}{1.4}
\begin{table}[bt]
  \caption{Maximum PSD limits in dBm/Hz for NB-PLC and BB-PLC.}
  \label{Tab:PSD_limit}
  \centering
    
\begin{threeparttable}
  	\begin{tabular}{|l|ccc|}
 \multicolumn{1}{c}{}  &  PS-OFDM & \shortstack{CB-FMT\\ HR} & \multicolumn{1}{c}{\shortstack{CB-FMT\\ LR }} \\ \hline \hline
     
     CENELEC A & -42.97 & -11.65 & -10.52  \\ \hline 
     CENELEC B & -55.76 & -21.1 & -21.74  \\ \hline 
     FCC-Low & -61.73 & -22.52 & -22.88  \\ \hline 
     FCC-Above & -81.11 & -72.66 & -70.89  \\ \hline 
     BB ($2$--$30$ MHz) & -56.62 & -55.72 & -55.12  \\ \hline 
   
  \end{tabular}
  
    The PSD limits are expressed in dBm/Hz.
 \end{threeparttable}

\end{table}

The table shows that PS-CB-FMT offers a PSD limit gain that ranges from $8.5$ to $39.5$ dB w.r.t. PS-OFDM in the HR configuration which can be increased even further with the LR configuration. This is due to the better spectrum confinement of the modulation scheme. That is, the spectrum falls down quickly outside the transmission band allowing to better meet the OOB limits. In turn, this enables to achieve higher spectral efficiency. 

\subsubsection{BB-PLC}
\begin{figure}[tb]
\centering
\includegraphics[width=\plotWidth]{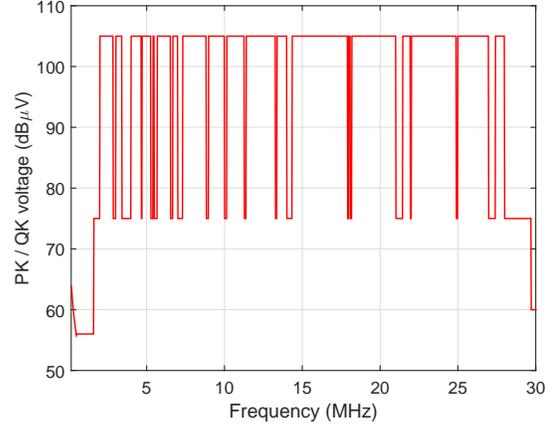}
   \caption{EMC mask for BB-PLC defined in the norm EN50561-1.}
   \label{Fig:mask}
\end{figure}

For BB-PLC, EN50561-1 \cite{EN50561_1} reports IB limits in terms of PK and AV values. Furthermore, spectrum notches are specified to allow coexistence with radio services. They are shown in Fig.~\ref{Fig:mask}. Although, EN50561-1 does not say how deep these notches have to be, in P1901 such notches are set $30$ dB below the limit in the transmission band, i.e., at $75$ dB$\mu$V (PK). It has also the be said, that notches are related to the regional EMC regulations, therefore P1901 allows flexibility on what sub-channels have to be switched on/off. Both PS-OFDM and PS-CB-FMT can transmit at maximum power over the active sub-channels and they switch off a number of sub-channels to meet the notching mask. Every notch in the mask is extended by 6 extra sub-carriers. 

\begin{figure}[tb]
\centering
\includegraphics[width=\plotWidth]{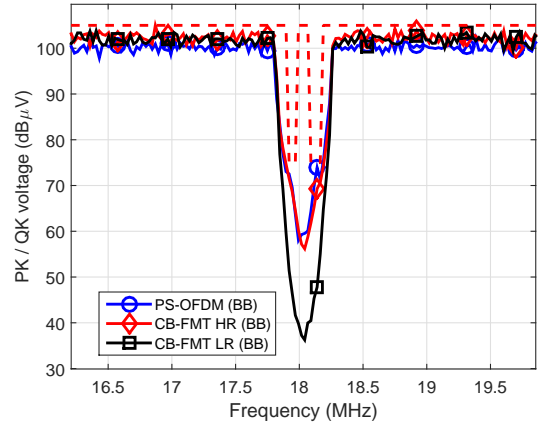}
   \caption{Zoom of EMC mask and spectra in the BB spectrum.}
   \label{Fig:bb_spectrum}
\end{figure}

Fig.~\ref{Fig:bb_spectrum} shows that PS-CB-FMT has a much better ability to create notches. In fact, both systems have the same active sub-channels. However, the notches in PS-CB-FMT are deeper and wider than those of PS-OFDM. This implies that the target notches can be fulfilled with a higher number of active sub-channels by PS-CB-FMT w.r.t. PS-OFDM.

\subsection{Maximum Achievable Rate}
\label{Sec:Results:Capacity}
To make a performance comparison between PS-OFDM and PS-CB-FMT, the maximum achievable rate is evaluated. The maximum achievable rate is computed in terms of Shannon capacity. 

\subsubsection{NB-PLC}
for NB-PLC, the O-LV application scenario is considered. O-LV involves the communication between the transformer sub-station and the houses. To model the channels, the OPERA model is considered \cite{OperaD5}. The model provides three classes of channels, as a function of the distance between the sub-station and the houses. For short (about 150 m) and long (about 350 m) distances the model provides three channel responses, namely good, medium and bad quality. For medium distance (about 250 m), only good and medium quality channels are provided. The model is derived from a measurement campaign.

For the background noise, OPERA also provides a model in \cite{OperaD5}. However, this model is developed for BB-PLC, i.e., for $f > 2$ MHz. For the lower frequencies the model is not defined. Thus, the model described in \cite{Outdoor_Noise} is considered. The noise is assumed to be stationary colored Gaussian. The PSD exhibits an exponential profile with higher values at lower frequency. The model describes the noise over the O-MV lines. Since the O-LV noise has a lower level than O-MV \cite{IEICE_PLC}, an offset of $-13$ dB is applied to the PSD. In detail, the following model is adopted:
\begin{equation}
	\mbox{PSD}(f) = a+be^{10^{-6}fc } \qquad \left[ \frac{\mbox{dBm}}{\mbox{Hz}} \right], \label{eq:noisePSD_NB}
\end{equation}
where $a=-106$, $b=52.98$ and $c = -0.0032$. 

\begin{figure}[tb]
\centering
\includegraphics[width=\plotWidth]{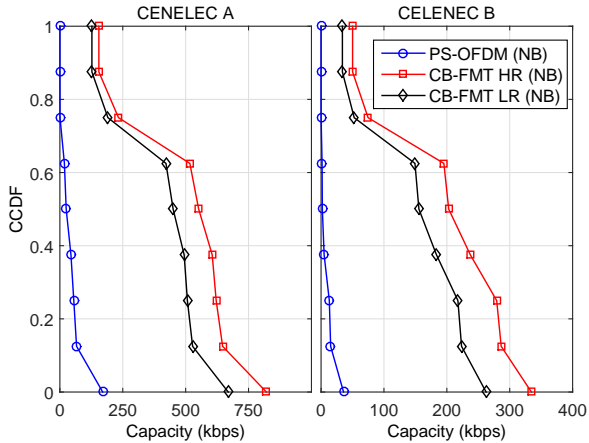}
   \caption{CCDF of the maximum achievable rate for the CENELEC bands.}
   \label{Fig:Capacity_CENELEC}
\end{figure} 

\begin{figure}[tb]
\centering
\includegraphics[width=\plotWidth]{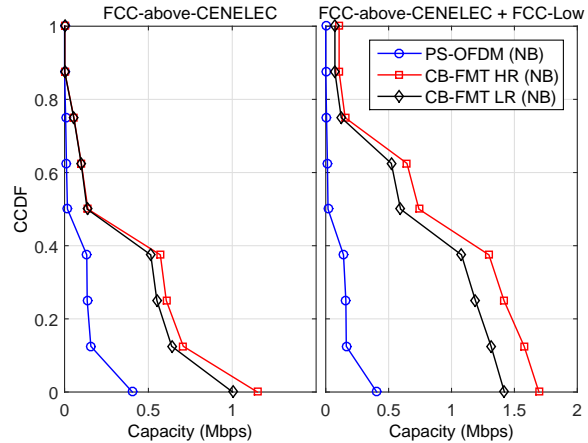}
   \caption{CCDF of the maximum achievable rate for the FCC bands.}
   \label{Fig:Capacity_FCC}
\end{figure}

Figs.~\ref{Fig:Capacity_CENELEC} and \ref{Fig:Capacity_FCC} show the Complementary Cumulative Distribution Function (CCDF) of the channel capacity in the CENELEC and FCC bands, respectively. PS-CB-FMT outperforms PS-OFDM both with the LR and HR configuration because the high frequency confinement allows to better match the norms and increase the power level (PSD limit) yet fulfilling the norms. Furthermore, PS-CB-FMT deploys matched filtering via frequency domain equalization at the receiver side which allows to better capture the signal energy \cite{Tonello_Girotto_JASP_2014}.
In CENELEC A, while PS-OFDM offers a capacity below 30 kbps with probability 0.5, CB-FMT in the HR configuration can reach up to 600 kbps. This shows that with CB-FMT there is great potentiality to increase performance.

\begin{figure}[tb]
\centering
\includegraphics[width=\plotWidth]{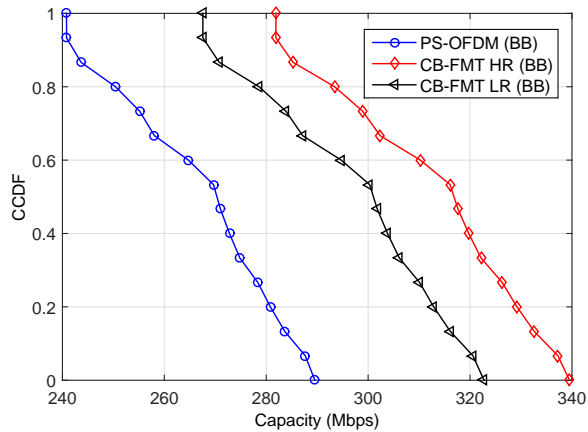}
   \caption{CCDF of the maximum achievable rate for BB-PLC.}
   \label{Fig:CCDF_BB}
\end{figure}
\subsubsection{BB-PLC}
for the IH scenario, the top-down model presented in \cite{Tonello_TopDown} is adopted. It accounts for the multipath propagation exhibited by the BB-PLC channel. The model parameters are set to match a large measurement campaign, described in \cite{TopDown_Omega,TopDown_measure}. Concerning the noise, the OPERA model is considered \cite{OperaD5}. The noise is modeled as additive, colored and Gaussian with the PSD according to \eqref{eq:noisePSD_NB} with $a=-107.625$, $b=28.694$ and $c = -0.044$. 

In Fig.~\ref{Fig:CCDF_BB}, the CCDF of the maximum achievable rate is reported. In general, PS-CB-FMT improves the performances w.r.t. PS-OFDM. In particular, with the LR configuration ($\alpha=496$, equal to PS-OFDM), the capacity improvement at CCDF 0.6 is from $268$ Mbps to $314$ Mbps. With the HR configuration, the performance improves by other  $15$ Mbps.   

\section{Conclusion}
\label{Sec:Conclusions}
In this work, a detailed overview of EMC norms relevant to both NB-PLC and BB-PLC has been reported. PLC is a particular communication technology that exploits the same port for both powering the modem and transmitting data. As such, it has required the development of new regulations which differ from continent to continent and that are still in progress. For instance, NB-PLC in the band $148.5$--$500$ kHz, as well as BB-PLC for outdoor applications, are unregulated in the EU. 

The measurement methodology has been analyzed to verify whether a PLC signal fulfills the EMC limits typically expressed in terms of PK, QP, or AV value. The procedure has been discussed aiming to provide a practical guide to the verification of the fulfillment of spectral constraints by a certain digital modulation system.   

In particular, PS-OFDM has been considered since it is the multicarrier modulation scheme used in state-of-the-art PLC systems. Using the IEEE P1901.2 specifications, it has been shown that PS-OFDM can fulfill the norms in the NB spectrum if the signal amplitude is significantly lowered w.r.t. what is prescribed by in-band EMC limits. This affects negatively the system capacity. Similarly, in the BB spectrum using IEEE P1901 specifications, a number of sub-channels must be switched off to create notches of $-30$ dB below the maximum PSD value. 

Therefore, an alternative and novel multicarrier scheme named PS-CB-FMT has been described and analyzed. Due to the higher spectrum confinement of PS-CB-FMT w.r.t. PS-OFDM, the spectral limits can be better achieved allowing for reduced out-of-band emissions and possibility to increase the in-band power up to the limit. Furthermore, PS-CB-FMT allows to switch off a lower number of sub-channels to create deep spectral notches. This translates into significantly higher system capacity offered by PS-CB-FMT w.r.t. PS-OFDM in both the NB and in the BB  spectrum.


\ifCLASSOPTIONcaptionsoff
  \newpage
\fi





\begin{thebibliography}{10}
\providecommand{\url}[1]{#1}
\csname url@samestyle\endcsname
\providecommand{\newblock}{\relax}
\providecommand{\bibinfo}[2]{#2}
\providecommand{\BIBentrySTDinterwordspacing}{\spaceskip=0pt\relax}
\providecommand{\BIBentryALTinterwordstretchfactor}{4}
\providecommand{\BIBentryALTinterwordspacing}{\spaceskip=\fontdimen2\font plus
\BIBentryALTinterwordstretchfactor\fontdimen3\font minus
  \fontdimen4\font\relax}
\providecommand{\BIBforeignlanguage}[2]{{%
\expandafter\ifx\csname l@#1\endcsname\relax
\typeout{** WARNING: IEEEtran.bst: No hyphenation pattern has been}%
\typeout{** loaded for the language `#1'. Using the pattern for}%
\typeout{** the default language instead.}%
\else
\language=\csname l@#1\endcsname
\fi
#2}}
\providecommand{\BIBdecl}{\relax}
\BIBdecl

\bibitem{PLC_state_art}
C.~Cano, A.~Pittolo, D.~Malone, L.~Lampe, A.~M. Tonello, and A.~G. Dabak,
  ``{State of the Art in Power Line Communications: From the Applications to
  the Medium},'' \emph{IEEE Journal on Selected Areas in Communications},
  vol.~34, no.~7, pp. 1935--1952, July 2016.

\bibitem{EN50065}
``{Signalling on Low-Voltage Electrical Installations in the Frequency Range 3
  kHz to 148,5 kHz - Part 1: General Requirements, Frequency Bands and
  Electromagnetic Disturbances},'' EN 50065-1:2011, CENELEC, 2011.

\bibitem{FCC_15}
``{Radio Frequency Devices},'' 47 CFR §15, FCC.

\bibitem{STD-T84}
``{Power line communication equipment (10kHz-450kHz)},'' STD-T84, ARIB, Nov
  2012.

\bibitem{MIMO_PLC_Book}
L.~Berger, A.~Schwager, P.~Pagani, and D.~Schneider, \emph{{MIMO Power Line
  Communications: Narrow and Broadband Standards, EMC, and Advanced
  Processing}}, ser. Devices, Circuits, and Systems.\hskip 1em plus 0.5em minus
  0.4em\relax Taylor \& Francis, 2014.

\bibitem{BookPLC}
L.~Lampe, A.~Tonello, and T.~Swart, \emph{{Power Line Communications:
  Principles, Standards and Applications from Multimedia to Smart Grid}}.\hskip
  1em plus 0.5em minus 0.4em\relax NY: Wiley \& Sons, 2016.

\bibitem{Bingham}
J.~A.~C. Bingham, ``{Multicarrier Modulation for Data Transmission, an Idea
  whose Time Has Come},'' \emph{{IEEE Communication Magazine}}, vol.~31, pp.
  5--14, May 1990.

\bibitem{Pulse-Shaping}
F.~Sjoberg, R.~Nilsson, M.~Isaksson, P.~Odling, and P.~O. Borjesson,
  ``{Asynchronous Zipper},'' in \emph{IEEE International Conference on
  Communications (ICC '99)}, vol.~1, Jun 1999, pp. 231--235.

\bibitem{Wavelet_OFDM}
S.~D. Sandberg and M.~A. Tzannes, ``{Overlapped Discrete Multitone Modulation
  for High Speed Copper Wire Communications},'' \emph{IEEE Journal on Selected
  Areas in Communications}, vol.~13, no.~9, pp. 1571--1585, Dec 1995.

\bibitem{OQAM}
B.~Saltzberg, ``{Performance of an Efficient Parallel Data Transmission
  System},'' \emph{IEEE Transactions on Communication Technology}, vol.~15,
  no.~6, pp. 805--811, December 1967.

\bibitem{OFDM_OQAM}
A.~Skrzypczak, P.~Siohan, and J.~P. Javaudin, ``{Application of the OFDM/OQAM
  Modulation to Power Line Communications},'' in \emph{IEEE International
  Symposium on Power Line Communications and Its Applications (ISPLC 2007)},
  March 2007, pp. 71--76.

\bibitem{FMT_Cherubini}
G.~Cherubini, E.~Eleftheriou, and S.~Olcer, ``{Filtered Multitone Modulation
  for Very High-Speed Digital Subscriber Lines},'' \emph{IEEE Journal on
  Selected Areas in Communications}, vol.~20, no.~5, pp. 1016--1028, Jun 2002.

\bibitem{FMT_Tonello_PLC}
A.~M. Tonello and F.~Pecile, ``{Efficient Architectures for Multiuser FMT
  Systems and Application to Power Line Communications},'' \emph{IEEE
  Transactions on Communications}, vol.~57, no.~5, pp. 1275--1279, May 2009.

\bibitem{Tonello_Girotto_JASP_2014}
A.~M. Tonello and M.~Girotto, ``{Cyclic Block Filtered Multitone Modulation},''
  \emph{EURASIP Journal on Advances in Signal Processing}, vol. 2014, no.~1, p.
  109, 2014.

\bibitem{CBFMT_GLOBECOM}
M.~Girotto and A.~M. Tonello, ``{Improved Spectrum Agility in Narrow-Band PLC
  with Cyclic Block FMT Modulation},'' in \emph{IEEE Global Communications
  Conference (GLOBECOM 2014)}, Dec 2014, pp. 2995--3000.

\bibitem{CBFMT_ISPLC}
A.~M. Tonello and M.~Girotto, ``{Cyclic Block FMT Modulation for Broadband
  Power Line Communications},'' in \emph{IEEE Int. Symposium on Power Line
  Communications and Its Applications (ISPLC 2013)}, March 2013, pp. 247--251.

\bibitem{EMC_Paul}
C.~Paul, \emph{{Introduction to Electromagnetic Compatibility}}, ser. Wiley
  Series in Microwave and Optical Engineering.\hskip 1em plus 0.5em minus
  0.4em\relax Wiley, 2006.

\bibitem{EMC_Engineering}
H.~Ott, \emph{{Electromagnetic Compatibility Engineering}}.\hskip 1em plus
  0.5em minus 0.4em\relax Wiley, 2011.

\bibitem{CISPR_22}
``{Information technology equipment -- Radio disturbance characteristics --
  Limits and methods of measurement},'' CISPR 22:2008, IEC, 2008.

\bibitem{EN55022}
``{Information Technology Equipment -- Radio disturbance characteristics --
  Limits and methods of measurement},'' EN 55022:2006, CENELEC, 2006.

\bibitem{CISPR_16_1}
``{Specification for radio disturbance and immunity measuring apparatus and
  methods -- Part 1: Radio disturbance and immunity measuring apparatus},''
  CISPR 16-1, IEC, 2002.

\bibitem{CISPR_I_89}
``{Amendment to CISPR 22: Clarification of its application to telecommunication
  system on the method of disturbance measurement at ports used for PLC (Power
  Line Communication)},'' CISPR/I/89/CD, IEC, 2003.

\bibitem{CISPR_I_301}
``{Amendment 1 to CISPR 22 Ed.6.0: Addition of limits and methods of
  measurement for conformance testing of power line telecommunication ports
  intended for the connection to the mains},'' CISPR/I/301/CD, IEC, 2009.

\bibitem{HomePlug_mask}
N.~Weling, ``{Expedient Permanent PSD Reduction Table as Mitigation Method to
  Protect Radio Services},'' in \emph{IEEE International Symposium on Power
  Line Communications and Its Applications (ISPLC 2011)}, April 2011, pp.
  305--310.

\bibitem{EN50561_1}
``{Power line communication apparatus used in low-voltage installations --
  Radio disturbance characteristics -- Limits and methods of measurement --
  Part 1 Apparatus for in-home use},'' EN 50561-1:2012, CENELEC, 2012.

\bibitem{cyclostationarity}
W.~A. Gardner, \emph{{Cyclostationarity in Communications and Signal
  Processing}}, ser. Electrical engineering, communications and signal
  processing.\hskip 1em plus 0.5em minus 0.4em\relax IEEE Press, 1994.

\bibitem{QP_digital}
F.~Krug and P.~Russer, ``{Quasi-Peak Detector Model for a Time-Domain
  Measurement System},'' \emph{IEEE Transactions on Electromagnetic
  Compatibility}, vol.~47, no.~2, pp. 320--326, May 2005.

\bibitem{oppenheim1999discrete}
A.~Oppenheim, \emph{Discrete-Time Signal Processing}, ser. Pearson education
  signal processing series.\hskip 1em plus 0.5em minus 0.4em\relax Pearson
  Education, 1999.

\bibitem{PRIME}
``{Narrowband Orthogonal Frequency Division Multiplexing Power Line
  Communication Transceivers for PRIME Networks},'' {Recommendation ITU-T
  G.9904}, 2012.

\bibitem{G3}
``{Narrowband Orthogonal Frequency Division Multiplexing Power Line
  Communication Transceivers for G3-PLC Networks},'' {Recommendation ITU-T
  G.9903}, 2012.

\bibitem{Ghnem_intro}
V.~Oksman and J.~Zhang, ``{G.HNEM: the new ITU-T standard on narrowband PLC
  technology},'' \emph{IEEE Communications Magazine}, vol.~49, no.~12, pp.
  36--44, December 2011.

\bibitem{GHNEM}
``{Narrowband Orthogonal Frequency Division Multiplexing Power Line
  Communication Transceivers for ITU-T G.hnem networks},'' {Recommendation
  ITU-T G.9902: Narrowband}, {ITU}, Oct. 2012.

\bibitem{P1901_2}
``{IEEE Standard for Low-Frequency (less than 500 kHz) Narrowband Power Line
  Communications for Smart Grid Applications},'' {IEEE 1901.2-2013}, Dec. 2013.

\bibitem{GHN}
``{Unified High-Speed Wireline-Based Home Networking transceivers - System
  Architecture and Physical Layer Specification },'' {Recommendation ITU-T
  G.9960.}, {ITU}, Oct. 2009.

\bibitem{P1901}
``{IEEE Standard for Broadband over Power Line Networks: Medium Access Control
  and Physical Layer Specifications},'' {IEEE 1901-2010}, Sep. 2010.

\bibitem{CFBM}
P.~P. Vaidyanathan and A.~Kirac, ``{Theory of Cyclic Filter Banks},'' in
  \emph{IEEE Int. Conf. on Acoustics, Speech, and Signal Processing
  (ICASSP-97)}, vol.~3, Apr 1997, pp. 2449--2452.

\bibitem{OperaD5}
M.~{Babic \textit{et al.}}, ``{D5 Pathloss as a function of frequency, distance
  and network topology for various LV and MV European powerline networks.}''
  IST Integrated Project No. 507667, Tech. Rep., April 2005, oPERA Deliverable
  D5.

\bibitem{Outdoor_Noise}
Z.~Tao, Y.~Xiaoxian, Z.~BaoHui, N.~Xu, F.~Xiaoqun, and L.~Changxin,
  ``{Statistical Analysis and Modeling of Noise on 10-kV Medium-Voltage Power
  Lines},'' \emph{IEEE Trans. on Power Delivery}, vol.~22, no.~3, pp.
  1433--1439, July 2007.

\bibitem{IEICE_PLC}
A.~M. Tonello, A.~Pittolo, and M.~Girotto, ``{Power Line Communications:
  Understanding the Channel for Physical Layer Evolution Based on Filter Bank
  Modulation},'' \emph{IEICE Transactions on Communications}, vol. E97-B,
  no.~8, pp. 1494--1503, August 2014.

\bibitem{Tonello_TopDown}
A.~M. Tonello, F.~Versolato, B.~Bejar, and S.~Zazo, ``{A Fitting Algorithm for
  Random Modeling the PLC Channel},'' \emph{IEEE Transactions on Power
  Delivery}, vol.~27, no.~3, pp. 1477--1484, July 2012.

\bibitem{TopDown_Omega}
M.~{Tlich \textit{et al.}}, ``{PLC Channel Characterization and Modelling},''
  Seventh Framework programme: Theme 3 ICT-213311 OMEGA, Tech. Rep., Dec 2008,
  deliverable D3.2.

\bibitem{TopDown_measure}
M.~Tlich, A.~Zeddam, F.~Moulin, and F.~Gauthier, ``{Indoor Power-Line
  Communications Channel Characterization Up to 100 MHz---Part I: One-Parameter
  Deterministic Model},'' \emph{IEEE Transactions on Power Delivery}, vol.~23,
  no.~3, pp. 1392--1401, July 2008.

\end{thebibliography}
\end{document}